\documentclass[12pt,preprint]{aastex}

\def\arcsecpoint{$''\!.$}
\def\deg{$^{\rm o}$}

\shortauthors{Crenshaw et al.}
\shorttitle{Geometry of Mrk~3}

\begin{document}

\title{The Geometry of Mass Outflows and Fueling Flows in the Seyfert~2
Galaxy Mrk~3\altaffilmark{1}}

\author{D.M. Crenshaw\altaffilmark{2},
S.B. Kraemer\altaffilmark{3},
H.R. Schmitt\altaffilmark{4},
Y.L. Jaff\'{e}\altaffilmark{5},
R.P. Deo\altaffilmark{6},
N.R. Collins\altaffilmark{7},
and T.C. Fischer\altaffilmark{2}}

\altaffiltext{1}{Based on observations made with the NASA/ESA Hubble Space 
Telescope, obtained at the Space Telescope Science Institute, which is 
operated by the Association of Universities for Research in Astronomy,
Inc. under NASA contract NAS 5-26555.}

\altaffiltext{2}{Department of Physics and Astronomy, Georgia State 
University, Astronomy Offices, One Park Place South SE, Suite 700,
Atlanta, GA 30303; crenshaw@chara.gsu.edu}

\altaffiltext{3}{Institute for Astrophysics and Computational Sciences,
Department of Physics, The Catholic University of America, Washington, DC
20064}

\altaffiltext{4}{Remote Sensing Division, Naval Research Laboratory,
Washington, DC 20375; and Interferometrics, Inc., Herndon, VA 20171}

\altaffiltext{5}{School of Physics and Astronomy, University of Nottingham,
University Park, Nottingham NG7 2RD, UK}

\altaffiltext{6}{Department of Physics, Drexel University, 3141 Chestnut
St., Philadelphia, PA 19104}

\altaffiltext{7}{Astrophysics Science Division, Code 667, Goddard Space
Flight Center, Greenbelt, MD 20771}

\begin{abstract}

We present a study of the resolved emission-line regions and an inner
dust/gas disk in the Seyfert 2 galaxy Mrk~3, based on {\it Hubble Space
Telescope} observations. We show that the extended narrow-line region
(ENLR), spanning $\sim$4 kpc, is defined by the intersection of the
ionizing bicone of radiation from the AGN and the inner disk, which is not
coplanar with the large-scale stellar disk. This intersection leads to
different position and opening angles of the ENLR compared to the
narrow-line region (NLR). A number of emission-line arcs in the ENLR appear
to be continuations of dust lanes in the disk, supporting this geometry.
The NLR, which consists of outflowing emission-line knots spanning the
central $\sim$650 pc, is in the shape of a backwards S. This shape
may arise from rotation of the gas, or it may trace the original fueling
flow close to the nucleus that was ionized after the AGN turned on.

\end{abstract}

~~~~~

\section{Introduction}

Mrk~3 (UGC~3426) has been classified as an SB0 (Adams 1977) or S0 galaxy
(Windhorst et al. 2002). It has been studied extensively in all wavebands,
because it harbors a bright active galactic nucleus (AGN). It is a Seyfert
2 galaxy, because its optical spectra show only narrow (FWHM $<$ 1000 km
s$^{-1}$) emission lines except in polarized-light spectra, which reveal
broad emission lines from a Seyfert 1-like nucleus (Schmidt \& Miller 1985;
Miller \& Goodrich 1990; Tran 1995). The central nucleus, which contains
the supermassive black hole (SMBH), accretion disk, and broad-line region
(BLR), is hidden from our direct line of sight by a large column ($N_H
\approx 1.3 \times 10^{24}$ cm$^{-2}$) of cold dusty gas (Pounds \& Page
2005; Bianchi et al. 2005), presumably in the shape of a torus (Antonucci
1993). At a redshift of $z = 0.013509$ based on H~I 21-cm radiation (Tifft
\& Cocke 1988), Mrk~3 is at a distance of $\sim$55 Mpc (for $H_0 = 73$ km
s$^{-1}$ Mpc$^{-1}$); at this distance, 1$''$ corresponds to a transverse
size of 270 pc.

Ground-based images of Mrk~3 in the $B$ and $I$ filters (Wagner 1987;
Schmitt \& Kinney 2000) show a host galaxy that is only slightly inclined,
with an ellipticity $e = 0.16$ (inclination $i = 33$\deg) and position
angle $P.A. = 28$\deg, based on its outer I-band isophotes at $r = 30'' -
40''$ from its nucleus. The ellipticity increases to $e = 0.36$ at $r =
10''$ at the same $P.A.$, suggesting the presence of a stellar bar and
supporting an SB0 classification.

{\it Hubble Space Telescope} ({\it HST}) images of Mrk~3 in the light of
[O~III] at an angular resolution of $\sim$0\arcsecpoint1 (Capetti et al.
1995, 1996, 1999; Schmitt \& Kinney 1996; Schmitt et al. 2003) reveal that
its narrow-line region (NLR) consists primarily of resolved emission-line
knots extending over $\sim$2\arcsecpoint4 ($\sim$650 pc), arranged in a
backwards S configuration resembling a grand-design spiral, similar to that
seen in several other Seyfert galaxies (Schmitt et al. 2003).
High-resolution radio observations (Kukula et al. 1993; 1999) show radio
knots along opposing jets that are nearly coincident with the linear
portion of the NLR, but at slightly different position angles ($P.A. =$
82\deg\ for the jets, 71\deg\ for the NLR). The jets are slightly curved at
their ends, in the direction of the offset emission-line knots
that define the S-like structure of the NLR, as shown in Figure 1 (from
Figure 4 in Kukula et al. 1999).

Outside of the NLR, emission-line images of Mrk~3 show an extended NLR
(ENLR) with an overall shape that resembles a projected bicone, first
noticed in ground-based observations (Pogge \& DeRobertis 1993), extending
over a range of $\sim$15$''$ ($\sim$4 kpc). Many Seyfert galaxies show
ENLRs, and their high ionization levels and often biconical shapes suggest
they are ionized by the central AGN. Their relative faintness and distinct
kinematics (dominated by rotation) compared to NLRs indicate that ENLRs
arise from ionization of the gas in galactic disks by the AGN (Unger et al.
1987). The above {\it HST} studies revealed that the ENLR in Mrk~3 shows
significant structure, which can be seen not only in the narrow-band
emission-line images but also in some of the broad-band images as well,
such as in the WFPC2 F606W image (Malkan, Gorjian, \& Tam 1998), due to
inclusion of strong emission lines such as [O~III] $\lambda\lambda$4959,
5007 and H$\alpha$ in the bandpass. The F606W image also includes a strong
contribution from the stellar continuum emission as well, which can be used
to identify dust structures in the circumnuclear regions.

\section{Previous Work and Open Questions}

In Figure 2, we show a structure map\footnote{A structure map is the
correction image obtained from the second iteration of a Richardson-Lucy
reconstruction. Starting with an appropriate {\it HST} point-spread
function (PSF), one divides the original image with a PSF-convolved version
of the image and multiplies the resulting image with the transpose of the
PSF. This procedure results in an image with the high-frequency component
enhanced in contrast (Pogge \& Martini 2002; Deo et al. 2006).}
of the F606W image of Mrk~3 that highlights the small-scale structure of
the emission-line regions and dust lanes, as well as the overall geometry
of the NLR and ENLR. As noted by others (Wagner 1987; Haniff, Wilson, \&
Ward 1988; Capetti et al. 1995), the ENLR, defined by the faint
emission-line arcs at greater distances from the nucleus than the S-shaped
NLR, appears to be rotated and spans a narrower opening angle than the NLR
(see also Figure 7). Thus, one question we would like to address in this
paper is: why are the position angle and opening angle of the ENLR
different from those of the NLR?

Figure 2 shows that much of the ionized gas in the ENLR of Mrk 3 is in the
form of curved structures, such as arcs or pieces of spirals. Furthermore
there appears to be a connection between some of these structures and the
dust lanes to the NE, similar to that seen in the NLR of the Seyfert 2
galaxy Mrk 573 (Quillen et al. 1999; Pogge \& Martini 2002; Martini et al.
2003a). Together, these features are similar in appearance to the nuclear
dust spirals found in most Seyfert galaxies (Regan \& Mulchaey 1999; Pogge
\& Martini 2002, Martini et al. 2003a), which likely trace the fueling
flows to the AGN in the inner few kiloparsecs (Martini et al. 2003b; Deo et
al. 2006; Sim\~{o}es Lopes et al. 2007). Another interesting question is
therefore: what is the nature of the structure in the ENLR, and how is it
related to the dust lanes that are outside of the ionizing bicone?

In Ruiz et al. (2001), we presented long-slit and slitless spectra of Mrk~3
obtained with the Space Telescope Imaging Spectrograph (STIS) on {\it HST},
which we used to study the kinematics of the emission-line knots in the NLR
based on the [O~III] $\lambda$5007 emission line. We found that the
spatially-resolved spectra could be matched with a biconical outflow model
that incorporates increasing velocity from near zero km s$^{-1}$ at the
nucleus (in the galaxy's rest frame) to $\sim$800 km s$^{-1}$ (in the line
of sight) at 0\arcsecpoint3 from the nucleus, followed by decreasing
velocity to near zero km s$^{-1}$ at an angular distance of
$\sim$1\arcsecpoint0. Our model adopted a rather narrow bicone (maximum
half-opening angle [$\theta_{max}$] $=$ 22.5\deg) with its axis positioned
along the central linear structure in Figure 2 ($P.A.$ $=$ 71\deg),
corresponding to our slit location (see Ruiz et al. 2001). However, this
bicone does not include the offset knots to the east and west that define
the S shape. Thus, we would like to re-examine this problem and address a
third question: what is the nature and origin of this shape, and can it be
reconciled with our general picture of radial outflow?

In a previous paper (Collins et al. 2005), we modeled the geometry of the
circumnuclear regions in Mrk 3. For the NLR bicone, we used the parameters
from our kinematic models of the outflow (Ruiz et al. 2001), which are the
$P.A.$ of the bicone axis, the inclination ($i$) of the bicone axis (zero
is defined to be in the plane of the sky), and the half-opening angle of
the outer edge of the bicone ($\theta_{max}$) . For the galactic disk we
used the $P.A.$ and $i$ from ground-based images of the galaxy (Schmitt \&
Kinney 1996). Based on the dust lanes seen in Figure 2, we suggested that
the east side of the disk is closer to us.

There are several problems with the above interpretation as it relates to
our open questions. First, the model cannot explain the observed ENLR. As
can be seen in Collins et al. (2005, Figure 20), the intersection between
the bicone and galactic disk defines the ENLR (seen clearly in the NE)
which has an axis of symmetry at smaller $P.A.$s than the NLR, the opposite
of what is observed. If we make the west side of the disk closer to us,
keeping the same values for the other parameters, the effect is the same,
because the major axis of the disk lies roughly along the NE-SW edge of the
bicone, and inclining the disk in either direction brings the E-W edge out
of the disk. The second problem is that the dust lanes are approximately
parallel to the minor axis of this disk, not the major axis as expected.
Finally, the bicone as originally defined is rather narrow, and does not
include the offset knots on either side of the inner, linear portion of the
NLR or the ENLR, which lies in the same direction as the offset knots.
Thus, we have undertaken a study to address these problems and the open
questions that we have posed. First, we describe a new analysis of existing
{\it HST} emission-line images to examine the ionization strucutre of the
NLR in Mrk~3. Then, we re-examine the geometry of the NLR, ENLR, and dust
structures in Mrk~3 in order to obtain a much better match to the
observational constraints. Finally, we discuss how our geometric model
provides new insights into the nature of fueling flows and outflows in this
Seyfert galaxy.

\section{Ionization Structure of the NLR and ENLR}

We generated an ionization map of the circumnuclear regions in Mrk~3 by
constructing an [O~III] $\lambda$5007/[O~II] $\lambda$3737 ratio image,
to probe the geometry and structure of these regions. The observations were
obtained with the post-COSTAR Faint Object Camera (FOC) on {\it HST} in
the f/151, 512$\times$512 mode, yielding a pixel size of 0\arcsecpoint014
(the angular resolution is $\sim$0\arcsecpoint1 FWHM). These observations
were made as part of the project GO-5140 (P.I. Macchetto) on 1994 March 20.
A log of the observations is presented in Table~1.

The images were retrieved from the {\it HST} archive, reduced, and
calibrated using standard procedures (see Schmitt \& Kinney 1996 for more
details). The images were aligned, background subtracted using regions free
of emission in the chip, and calibrated using information available in the
header. The continuum image was used to subtract the host galaxy
contribution to the emission-line images. This was done by scaling the flux
of the continuum image based on the width of the two filters. We used the
same continuum image for both on-line images. Although it would have been
better to use a bluer continuum image for the subtraction of the host
galaxy contribution to the [O~II] image, the only continuum image that was
not strongly contaminated by line emission was that obtained with the F550M
filter. This choice should not significantly affect the resulting
continuum-subtracted [O~II] image. The host galaxy of Mrk~3 is an SB0, so
we do not expect large color variations along the galaxy or large amounts
of dust. We checked on whether this procedure over- or under-subtracted the
host galaxy by inspecting the outer regions of the pure emission-line
images, and further small corrections were applied when neccesary to reduce
the residuals. We estimate that the uncertainty in the continuum
subtraction is on the order of 5\%.

The final steps of the data reduction involved the determination of the
noise in regions free from emission, used to determine the uncertainty in
the flux measurements. Based on the [O~II] image, the one with the highest
noise, we created a mask that blanked all regions with flux below the
3$\sigma$ level. This mask was applied to both images, to ensure that we
were not comparing regions with strong emission in one image with noise in
the other. The blanked [O~II] and [O~III] images were used to measure the
emission line fluxes in different regions of the NLR. This was done using
15 concentric circular annuli, 8 pixels ($\sim0.1$\arcsec) wide. Each
annulus was split into 20 sectors, with widths of 18$^{\circ}$. Sectors
with more than 50\% of the [O~II] or [O~III] pixels blanked were eliminated
from our analysis.

In Figure 3, we show the [O~III]/[O~II] ratio image, and in Figure 4, we
show plots of the [O~III]/[O~II] ratio as function of $P.A.$ at various
distances from the central nucleus. Outside of the central nucleus, the
ratio peaks at values of 4 -- 6 around position angles $\sim$100\deg\ and
$\sim$240\deg, at the locations of bright emission-line knots across the
NLR, consistent with the STIS long-slit values in Collins et al. (2005).
These ratios hold for the eastern offset knot and the core of the western
offset knots, indicating ionization levels similar to those in the inner,
linear portion of the NLR. The more diffuse emission surrounding the
western knot and the northern edge of the NLR have somewhat lower values,
indicating lower ionization parameters. These results are consistent with
the suggestion by Collins et al. (2009) that there is lower ionization gas
along the edges of the bicone that is due to filtering of the ionizing
radiation by an absorber that is close to the central source. The
[O~III]/[O~II] ratios in the ENLR (right-side plots in Figure 4) are lower,
although still AGN-like, which could be due to filtering of the ionizing
radiation as it traverses the ENLR disk.

The high ionization levels and high emissivities of the offset knots
demonstrates that they see the ionizing flux directly and are therefore not
``fossil nebulae'', in which the ionizing radiation has been cut off
(Binette \& Robinson 1987). Thus, the offset knots are in the ionizing
bicone, which must be much wider than we previously assumed. Furthermore,
the differences in $P.A.$ between the central NLR versus the offset knots
and ENLR cannot be explained by a change in the $P.A.$ of a narrow bicone
of ionizing radiation, due, for example, to precession of an
optically-thick torus. The light-travel time to the ends of the linear
portion of the NLR is $\sim$1000 yr, which indicates that the bicone has
maintained its current position for at least this long. If we assume that
the jets are traveling at $\sim$0.1c (Ulvestad 2005, and references
therein), the linear structure out to large distances indicates they have
maintained their positions for even longer, $\sim$10,000 yr. The
recombination time for [O~III] is $t_{rec} \approx [\alpha($O$^{+2},T)
~n_e$]$^{-1}$, where the recombination coefficient $\alpha$(O$^{+2},T$) $=$
1.72 $\times$ 10$^{-11}$ cm$^3$ s$^{-1}$ at $T = 10,000$K (Osterbrock \&
Ferland 2006). After $\sim$1000 yr or longer, the O$^{+2}$ in the offset
knots and ENLR would have recombined, unless the electron densities are
$n_e < 2$ cm$^{-3}$. This is not feasible, because such low densities would
lead to very low emissivities and the emission-line knots would not be
visible. In fact, according to the trend of decreasing density from the
nucleus from our studies of the STIS long-slit spectra of Mrk~3 (Collins et
al. 2009), the offset knots should have a density of $n_e \approx 1000$
cm$^{-3}$, yielding a recombination time of only $\sim$2 yr for O$^{+2}$.
Thus, precession is not a valid explanation for the different $P.A.$s of
the NLR and ENLR. Finally,if one supposed that the difference was due to
gas in the ENLR rotating out of a narrow bicone, it would take
$\sim$10$^{6}$ yr, assuming a galactic rotational velocity of 250 km
s$^{-1}$, so this is not a valid explanation either. Thus, we must look for
another way to explain the different position and opening angles of the NLR
and ENLR.

\section{Modeling the Geometry of the NLR and ENLR}

We have taken a fresh approach to the geometry of the NLR, inner disk, and
ENLR. We remeasured the NLR bicone to include both the linear structure and
offset knots, which results in a much wider bicone at a somewhat larger
$P.A.$, but with the same inclination angle. We take the apex of the bicone
to be the location of the hidden nucleus established by ultraviolet imaging
polarimetry (Kishimoto, et al. 2002), which is consistent with the location
of the optical continuum peak and kinematic center of the NLR (Ruiz et al.
2001; Collins et al. 2005). We give our NLR bicone parameters, as well as
our measurements of the projected ENLR, in Table 2. We determined an
average $P.A.$ for the dust lanes of 129\deg ($\pm$6\deg), and assumed this
was the $P.A.$ of an inner dust/gas disk on the same scale as the ENLR. We
again took the NE side of the disk to be the closer side, based on the
presence of dust lanes only on this side. This is also consistent with
our finding that the NLR emission is more reddened in the east than in the
west (Collins et al. 2005). The only parameter that we did
not determine directly was the inclination angle of the inner disk.

We have developed a geometric modeling program, similar to that used by
Mulchaey et al. (1996), to visualize the intersection between the inner
dust/gas disk and the bicone of ionizing radiation. We varied the
inclination angle of the disk, which changes the intersection between the
disk and bicone of ionizing radiation, to match the $P.A.$ and
$\theta_{max}$ of the ENLR. We were able to match the measured values to
within the measurement uncertainties ($\pm$ 2\deg) with $i$ $=$ 64\deg
$\pm$2\deg. Figure 5 shows the resulting geometric model from our
viewpoint, and Figure 6 displays a view down the eastern cone. In Figure 5,
the solid portion on the eastern side lies directly above the intersection
between the eastern cone and disk -- the ENLR is an extension of this
triangle plus a symmetric triangle in the other direction, as shown in
Figure 7. As shown in Figure 6, the intersection between the bicone and
disk is in the southern part of the east cone, resulting in the smaller
opening angle and shifted $P.A.$ of the ENLR compared to the NLR bicone as
seen from the Earth.

Figure 7 shows the structure map from Figure 2, with the outlines of our
geometric model superimposed. The model encompasses nearly all of the
emission-line structure in both the NLR and ENLR, except for faint
extentions of a couple of arcs in the ENLR. These can be explained by
evidence in this paper and in Collins et al. (2009) that the bicone is not
sharp-edged in Mrk~3, and this gas may in fact be ionized by weak, filtered
radiation from the AGN at polar angles slightly greater than
$\theta_{max}$.

\section{Conclusions}

We have developed a new geometric model for the NLR and ENLR in the Seyfert
2 galaxy Mrk~3 that eliminates the discrepancies that we have encountered
with our previous model (Ruiz et al. 2001; Collins et al. 2005), explains a
number of observed properties of these regions, and helps to elucidate the
geometry and nature of the fueling flows traced by dust structures and
outflows traced by ionized gas kinematics in the circumnuclear
regions. Our new model has a much larger opening angle and a slightly
different $P.A.$ for the radiation bicone, to explain the ionization of not
only the inner, linear portion of the NLR, but the offset emission-line
knots and the entire ENLR as well. We previously assumed an inner dust/gas
disk that was at the same $P.A.$ and $i$ as the disk of the host galaxy,
but that lead to two problems: 1) the dust lanes are parallel to the minor
axis of the disk, and 2) the intersection between the disk and radiation
bicone cannot explain the differences between the NLR and ENLR in $P.A.$
and opening angle. However, by adopting an inner dust/gas disk that is
parallel to the observed inner dust lanes and varying the inclination of
the disk, we were able to match the geometry of the NLR and ENLR and
explain these differences, answering our first open question.

Evidence for an inner dust/gas disk in Mrk 3 comes not only from the dust
lanes to the NE of the nucleus, but from the emission-line arcs in the ENLR
as well. In fact, it appears that in some cases, the emission-line arcs are
continuations of the dust lanes into the ENLR, with the associated gas
ionized by the AGN, supporting our claim that the geometry of the ENLR is
due to the intersection between the dust/gas  disk and the radiation bicone
defined by the NLR, answering our second open question. This is the same
scheme that has been proposed for the emission-line arcs in Mrk~573
(Quillen et al. 1999; Martini et al. 2003a; Schlesinger et al. 2009).

Where does the inner dust/gas disk come from? Noordermeer et al. (2005)
discovered a bridge of H I emission between Mrk~3 and UGC~3422, a companion
spiral galaxy $\sim$100 kpc to the NW. They suggest that this gas, which
shows a local concentration in Mrk~3, was tidally drawn out of the gas
disk in UGC~3422. The bridge of gas is in the same general direction as the
major axis of the inner disk in Mrk~3, which is at a $P.A. = 129$\deg (or
$-$51\deg). This gas is therefore the likely source of the inner disk and
the large-scale fueling flow to the AGN.

\section{Discussion}

We have shown how our new geometric model of Mrk~3 explains the different
$P.A.$s and opening angles of its NLR and ENLR, as well as the arc-like
structure of the emission-line gas in the ENLR. The remaining question to
be addressed is the nature of the backwards S shape of the NLR. This
structure is not unique to Mrk~3, and may therefore be of some fundamental
importance. From a study of the NLRs in 60 Seyfert galaxies using {\it HST}
[O~III] emission-line images (Schmitt et al. 2003), three others show
well-defined S (or Z) shapes: NGC~3393 (Seyfert 2), NGC 3516 (Seyfert 1),
and NGC 6860 (Seyfert 1). Several other NLRs show curved structures or
arcs, like those in Mrk 573 or in the ENLR of Mrk~3.

The offset knots that define the backwards S shape in Mrk~3 appear to be
well separated from the inner, linear portion of the NLR, and lie along the
same general direction as the ENLR. Thus, they could lie in the same disk
as the ENLR, but their other properties are more like the NLR knots. They
have higher emissivities and higher velocity dispersions than the ENLR
knots (Ruiz et al. 2001), and their [O~III]/[O~II] ratios are similar to
the other NLR knots; these all suggest that they belong to the NLR. Because
the offset knots lie in the same direction as the ENLR, they could possibly
be located in the inner dust/gas disk. However, it is not clear where the
linear portion of the NLR is located within the bicone. According to our
geometry, the linear portion cannot be in the inner disk, unless the disk
begins to warp inside of the offset knots.

The kinematics of the NLR in Mrk~3 (Ruiz et al. 2001) follow the same
pattern seen in the other three Seyfert galaxies that we have studied in
detail using STIS long-slit spectra: NGC~4151 (Das et al. 2005), NGC 1068
(Das et al. 2006), and Mrk~573 (T. Fischer, et al., in preparation). The
emission-line knots have their own peculiar velocities superimposed on a
general pattern of radial outflow. So how does the backwards S shape of the
NLR in Mrk~3 fit into our picture of radial outflow? If all of the
outflowing gas in the NLR came from close to the nucleus (e.g., inside of
$\sim$30 pc, a resolution element), then how did this shape come about? It
could result from rotation of the outflowing gas, due to its initial
angular momentum, or it could come from collision of the gas with the
rotating inner disk. An advantage of the latter is that it might explain
the curved ends of the radio jets (Kukula et al. 1993). Unfortunately, we
cannot test for rotation, because we only have one STIS long-slit
observation, which does not pass through the offset NLR knots. The STIS
slitless data are much less precise; they indicate that the eastern offset
knot is blueshifted by $\sim$250 km$^{-1}$ and the western offset knot is
blueshifted by $\sim$50 km$^{-1}$ (Ruiz et al. 2001), but the latter is
uncertain by at least 50 km s$^{-1}$ due to the diffuseness of this knot.
STIS long-slit spectra of the offset knots, preferably at high spectral
resolution, would be extremely useful for testing the rotation hypothesis.

Another possible explanation for the backwards S shape is that it was
already present when the AGN turned on, after which at least some of the
gas was ionized. One intriguing possibility is that this shape traces the
original fueling flow to the nucleus on the scale of the NLR. This
suggestion is motivated by the fact that this shape resembles a
``grand-design'' nuclear dust spiral, which has been seen on the same scale
as the NLR in other Seyfert galaxies, along with other types of dust
spirals (Martini et al. 2003a), and offers an efficient way to fuel the
inner nucleus (Deo et al. 2006). This explanation is not inconsistent with
radial outflow, since the time required to disrupt the S shape is rather
long, on the order of the NLR crossing time, which is $\sim$5 $\times$
10$^{5}$ yr for a distance of 300~pc and a typical NLR cloud velocity of
500 km s$^{-1}$.

The above explanation would imply that much of the outflow that we see in
the NLR is due to in situ acceleration, rather than acceleration of clouds
from close to the nucleus. Support for in situ acceleration comes from our
determination that mass outflow occurs throughout the NLR in Mrk~3, and
that the mass of the NLR is $\sim$1 $\times$ 10$^{7}$ M$_{\odot}$ (Collins
et al. 2009). It seems unlikely that such a large amount of gas could
originate from inside of the unresolved nucleus, with a transverse size
$\leq$ 27 pc. Furthermore, we found that the mass outflow rate in the NLR
is $\sim$15 M$_{\odot}$ yr$^{-1}$, compared to the accretion rate of only
0.35 M$_{\odot}$ yr$^{-1}$ needed to sustain its bolometric luminosity
(Collins et al. 2009), and it is difficult to understand how all of this
outflow could come from close in. However, in situ acceleration presents a
couple puzzles. First, how does in situ acceleration explain the overall
flow pattern in the NLR, in which velocity increases and then decreases
with distance from the central AGN (Ruiz et al. 2001)? Second, why do the
NLR clouds have low dust/gas ratios and relatively low optical depths
(Collins et al. 2009), if they are accelerated off the cool inflowing gas?

A possible explanation is that the clouds are lifted off the fueling flow
by a highly ionized wind, perhaps in the form of a thermal or Parker wind
(Parker 1965). As suggested by Everett \& Murray (2007), the velocities of
the clouds are therefore controlled by the flow pattern of the wind. If
the wind originates inside the dust sublimation radius, then it will be
dust free, and mixing with the cooler gas in the clouds would lead to lower
dust/gas ratios. Alternatively, the low dust/gas ratios could result from
grain destruction during the interaction between the wind and the clouds.
The wind modeled by Everett \& Murray is more highly ionized than the X-ray
emission line gas that is extended along the optical NLR in Mrk 3 (Sako et
al. 2000; Bianchi et al. 2006), and it may have characteristics more in
common with the material that scatters the hidden BLR emission into our
line of sight. The major problem with this model is that the thermal wind
likely accelerates too rapidly to account for the radial velocity profile
observed in Seyfert NLRs (Everett \& Murray 2007). Furthermore, it is
unclear how a highly ionized wind could entrain the clouds, when the
outward pressure, due to radiation, on the latter could be much greater
than that on the wind. An alternative possibility is that the clouds are
radiatively driven, but their acceleration is inefficient due to their
interaction with the highly-ionized gas. The interaction is likely to
produce Kelvin-Helmholtz instabilities, which may cause microturbulence
(Kraemer, Bottorff, \& Crenshaw 2007), resulting in the large observed
velocity dispersions (Ruiz et al. 2005).

In any case, it is clear that the shapes of the NLR and ENLR in Mrk~3 are
defined by the intersection of the ionizing bicone of radiation and the
location of the gas. If the fueling flow to the AGN is continuous and has
not been disrupted by the outflow, the inner dust/gas disk must eventually
warp to $P.A. \approx$ 0\deg\ and $i =$ 5\deg\ to form the putative
optically-thick torus that produces the bicone of ionizing radiation in the
Mrk~3. Observations of cold molecular gas at high angular resolution would
help to identify possible warping and further elucidate the interaction
between fueling flows and outflows in Mrk~3.

\acknowledgments

Some of the data presented in this paper were obtained from the
Multimission Archive at the Space Telescope Science Institute (MAST).
Support for this work was provided by NASA through grant number
HST-AR-11243.04-A from the Space Telescope Science Institute, which is
operated by AURA, Inc., under NASA contract NAS 5-26555.

\begin{deluxetable}{lcll}
\tablecaption{FOC Observing Log\label{tb-1}}
\tablewidth{0pt}
\tablehead{
\colhead{Dataset Name}&\colhead{Exp.
Time}&\colhead{Filter}&\colhead{Comment}\\
\colhead{}&\colhead{(s)}&\colhead{}&\colhead{}}
\startdata
X2580103T &  750 & F502M & [O III]$\lambda$5007\\
X2580104T & 1196 & F550M & Continuum\\
X2580102T &  896 & F372M & [O II]$\lambda$3727\\
\enddata
\end{deluxetable}

\begin{deluxetable}{lll}
\tablecolumns{3}
\footnotesize
\tablecaption{Geometric Parameters for Mrk~3$^a$}
\tablewidth{0pt}
\tablehead{
\colhead{Parameter} & \colhead{Old Values$^b$} & \colhead{New Values$^b$}
}
\startdata
$P.A.$ (Disk)               &28\deg      &129\deg      \\
$i$ (Disk)                  &33\deg (NE) &64\deg (NE)  \\
$P.A.$ (NLR Bicone)         &71\deg      &89\deg       \\
$i$ (NLR Bicone)            &5\deg (NE)  &5\deg (NE) \\
$\theta_{max}$ (NLR Bicone) &25\deg      &51\deg       \\
$P.A.$ (ENLR)               &---         &112\deg      \\
$\theta_{max}$ (ENLR)       &---         &28\deg      \\
\enddata
\tablenotetext{a}{The letters in parentheses indicates the side closest to
us.}
\tablenotetext{b}{Old values are for the outer stellar disk. New values
are for the inner dust/gas disk.}
\end{deluxetable}

\clearpage

\figcaption[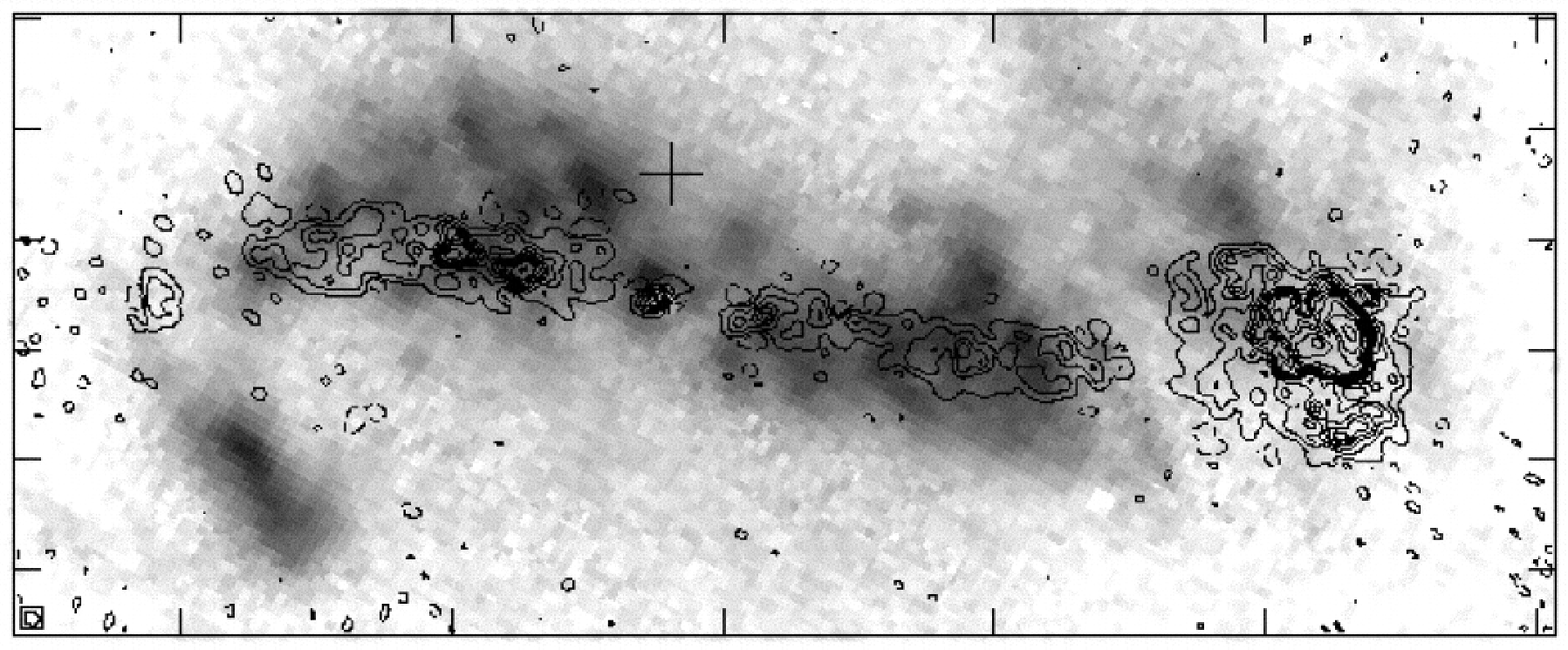]{[O~III] image (greyscale) of the NLR in Mrk~3 from the
{\it HST} Faint Object Camera, with radio 18 cm emission (contours)
superimposed, from Kukula et al. (1999, reproduced by permission of the
authors and the AAS). Tick marks on the vertical axis are separated by
$\sim$0\arcsecpoint2. North is up and east is to the left for this image
and all others.}

\figcaption[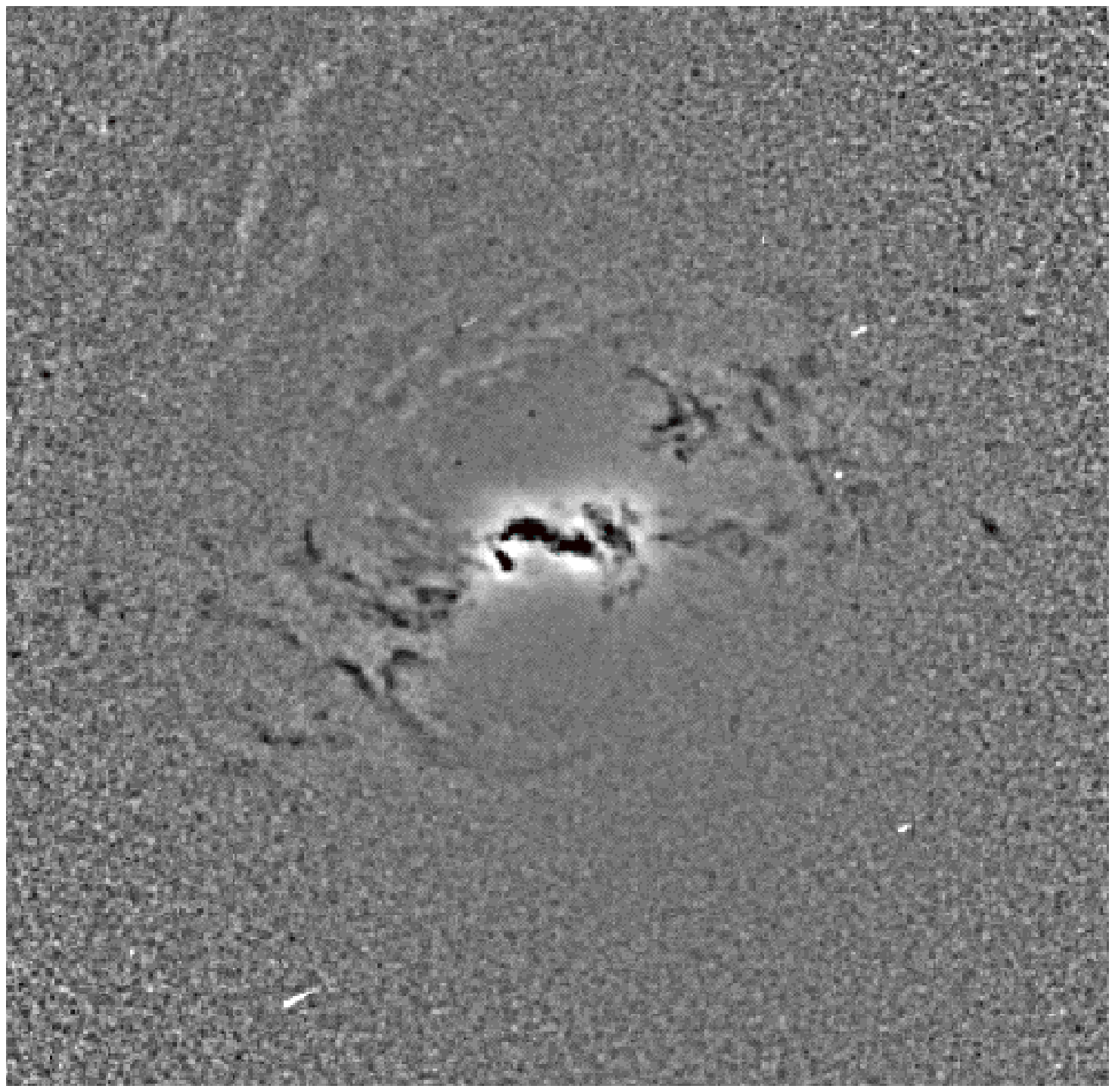]{Structure map of the {\it HST} WFPC2 image of Mrk~3
obtained with the F606W filter. The image is 23$''$ $\times$ 23$''$. Dark
areas correspond to line emission and bright areas correspond to dust
absorption. The white areas around the NLR are artifacts of the
structure-map procedure.}

\figcaption[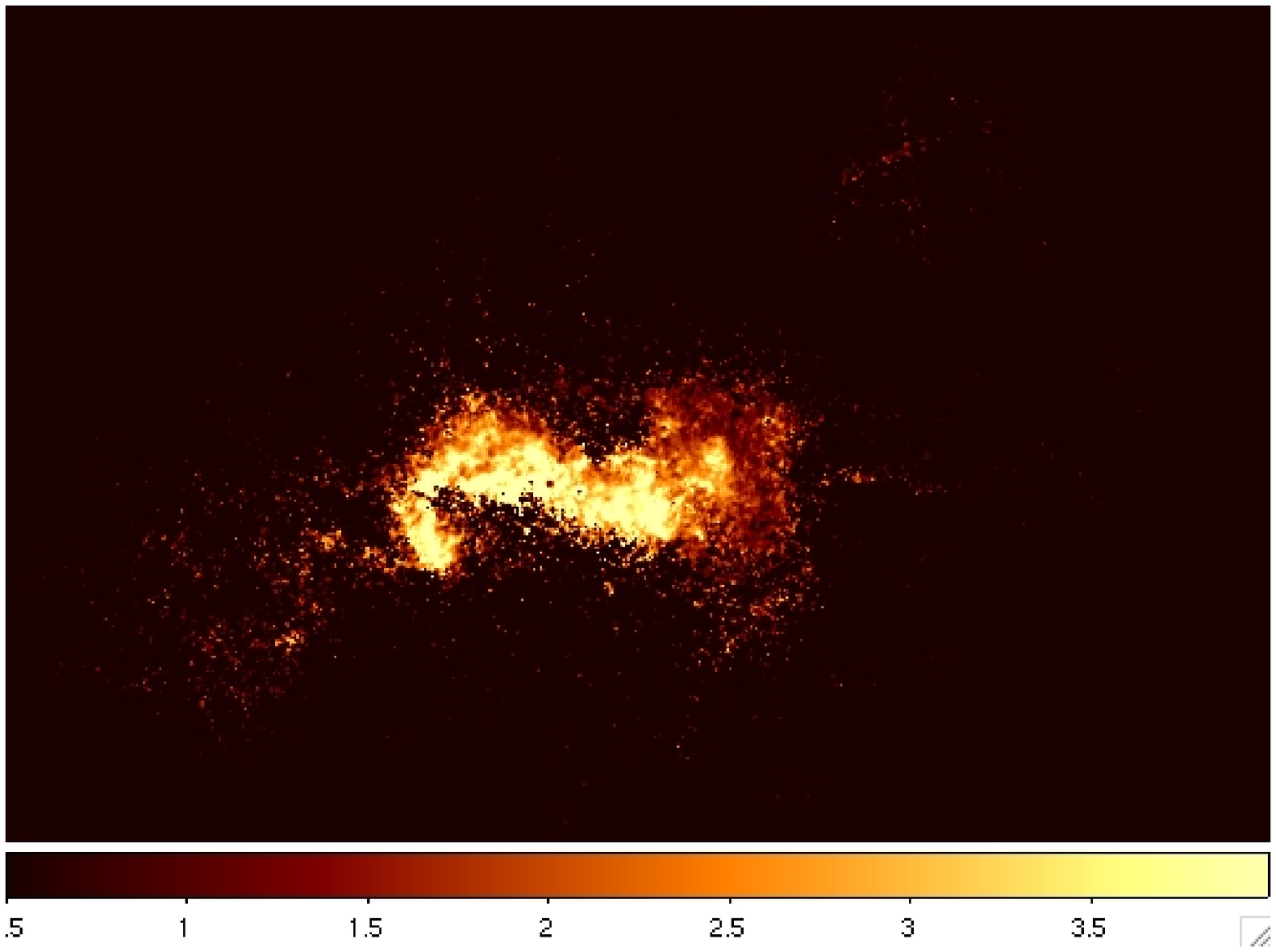]{[O~III]/[O~II] ratio map for the NLR in Mrk~3, based
on {\it HST} images. The color bar below the image gives the range of
[O~III]/[O~II] values.}

\figcaption[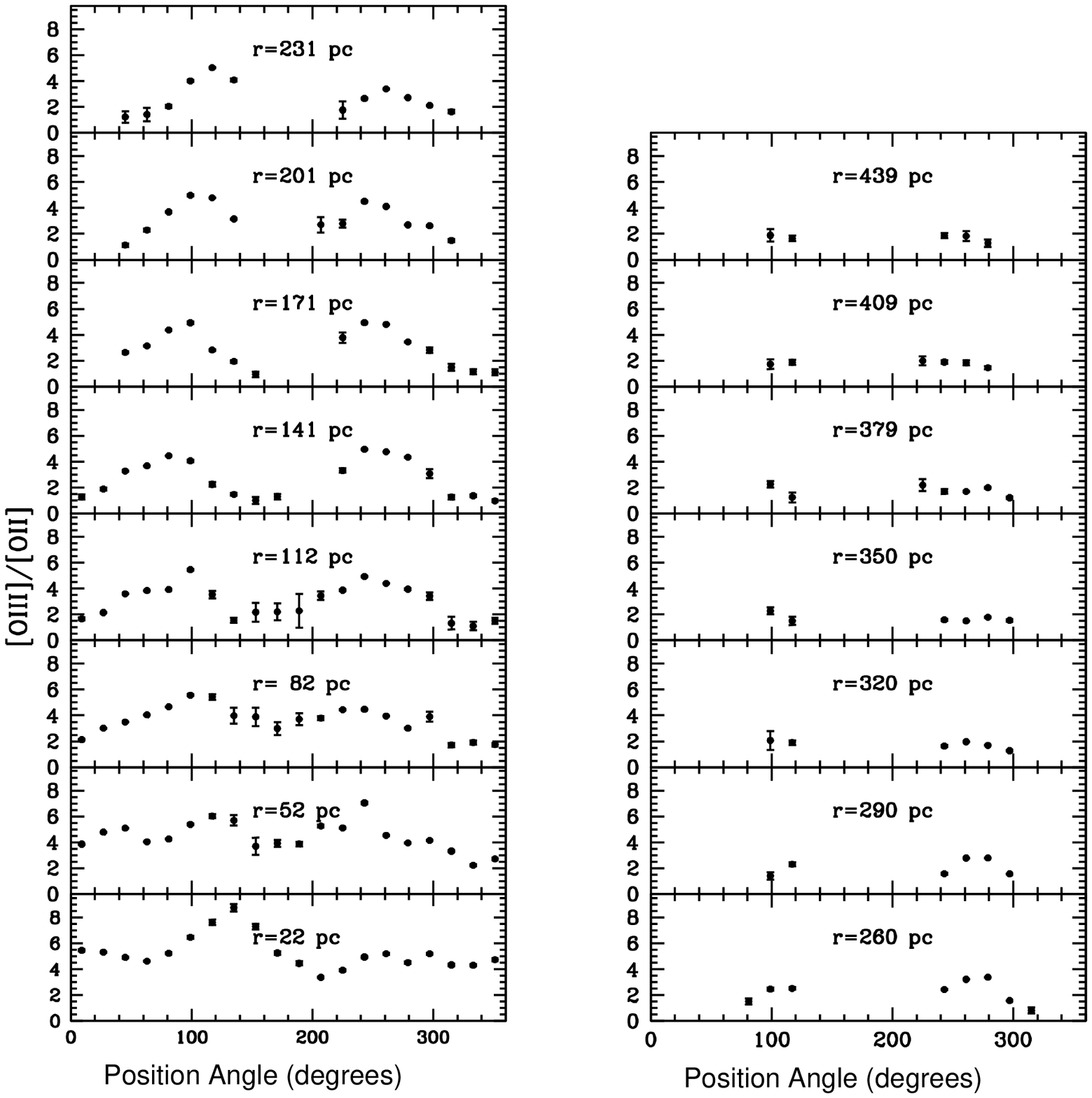]{[O~III]/[O~II] ratio as function of $P.A.$ at various
distances from the central nucleus. Values for the NLR and ENLR are shown
in the left and right plots, respectively.}

\figcaption[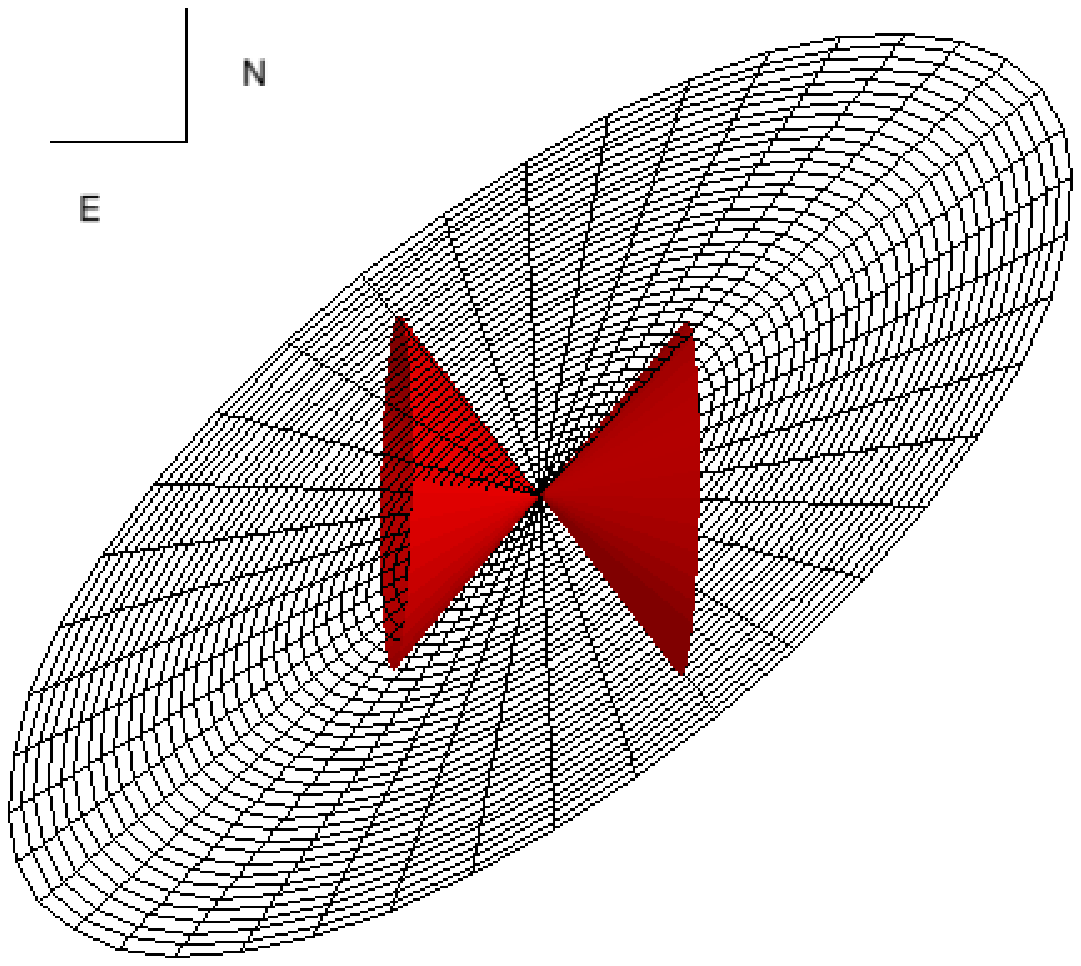]{Geometric model of the NLR bicone and inner disk in
Mrk~3, based on the ``new'' parameters in Table 2. This figure
shows the view from Earth. The NLR bicone axis is nearly in the plane of
the sky (the eastern axis lies 5\deg\ out of the plane). The NE side of the
disk is closer than the SW side and lies in front of most of the eastern
cone.}

\figcaption[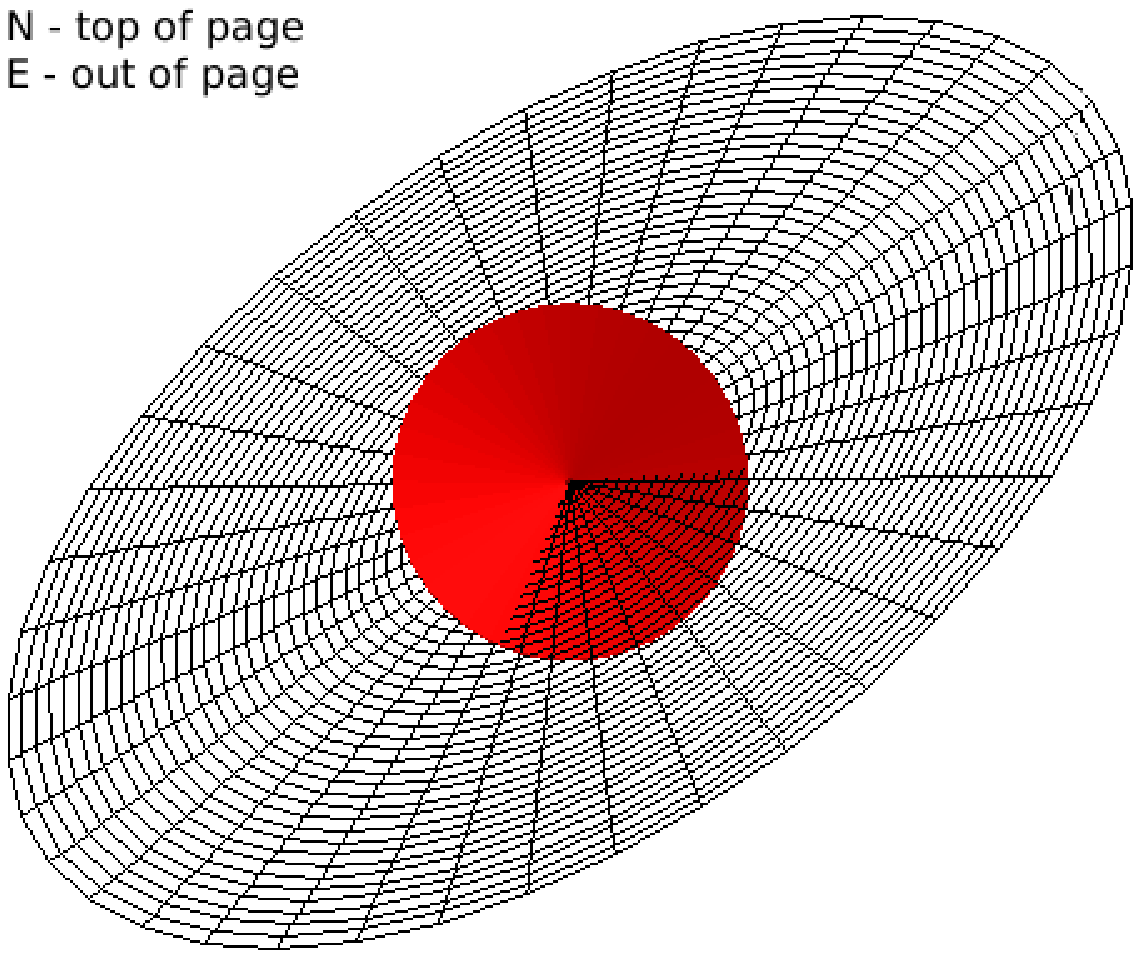]{Geometric model of the NLR bicone and inner disk in
Mrk~3, based on the ``new'' parameters in Table 2. This figure shows a view
down the eastern cone, with the view from Earth coming from the right, east
almost directly out of the page, and north at the top.}

\figcaption[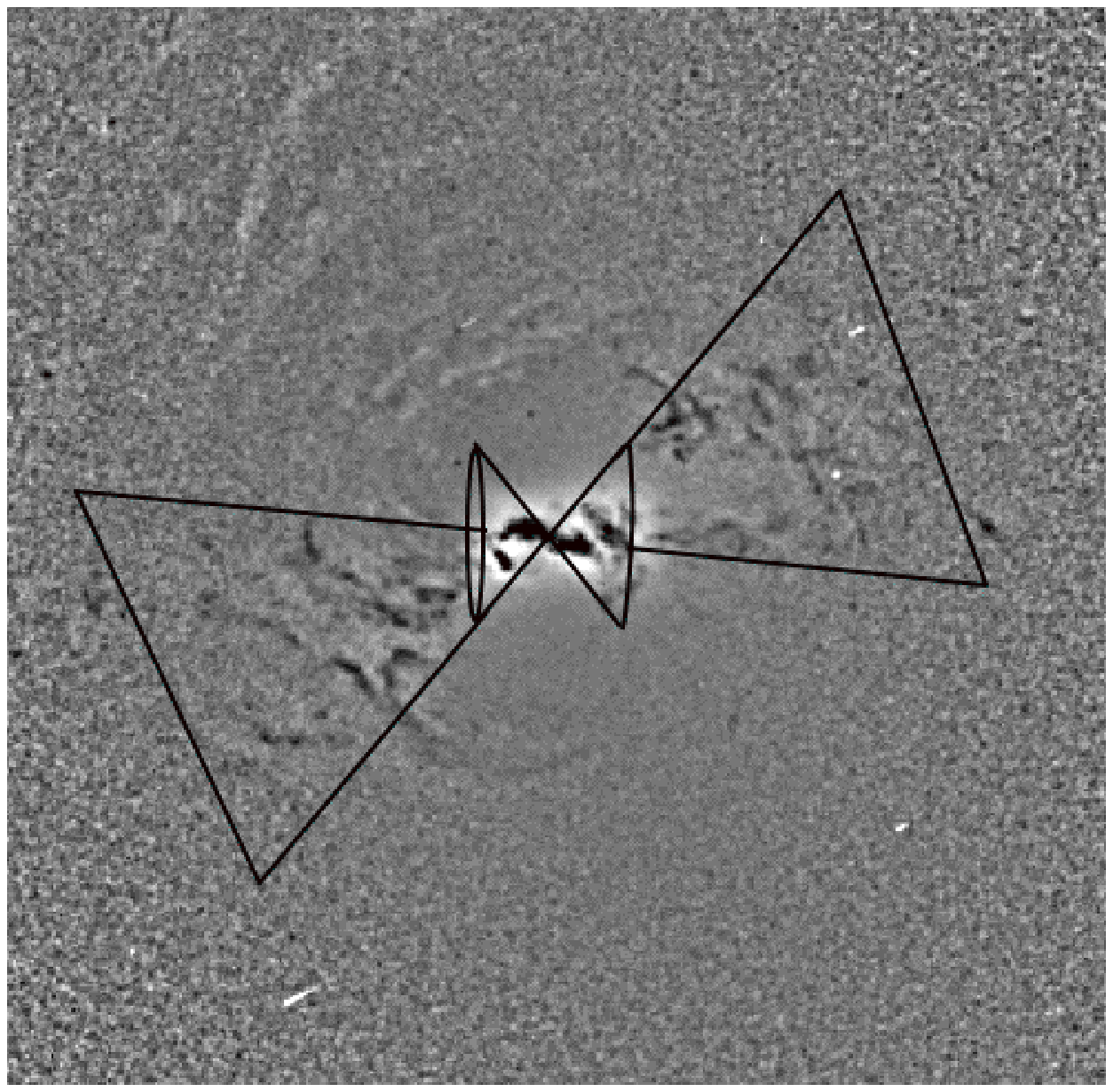]{Structure map of Mrk~3 with our model superimposed.
We show the inner bicone of ionizing radiation, which encompasses the
backwards S shape of the NLR. The outer triangular regions show the
intersection between the bicone of ionizing radiation and the inner dusty
disk, which encompasses the ENLR.}

\clearpage
\begin{figure}
\plotone{f1.eps}
\\Fig.~1.
\end{figure}

\begin{figure}
\plotone{f2.eps}
\\Fig.~2.
\end{figure}

\begin{figure}
\plotone{f3.eps}
\\Fig.~3.
\end{figure}

\begin{figure}
\plotone{f4.eps}
\\Fig.~4.
\end{figure}

\begin{figure}
\plotone{f5.eps}
\\Fig.~5.
\end{figure}

\begin{figure}
\plotone{f6.eps}
\\Fig.~6.
\end{figure}

\begin{figure}
\plotone{f7.eps}
\\Fig.~7.
\end{figure}

\end{document}